\documentclass{revtex4}
\usepackage{graphicx}
\usepackage{amsmath}
\usepackage{amsfonts}
\usepackage{graphics,graphicx}
\usepackage{color}

\newcommand{\emma}[1]{{#1}}
\newcommand{\gta}[1]{{#1}}
\newcommand{\gtad}[1]{{#1}}

\begin{document}
\title{Like-charge attraction in a one-dimensional setting:
the importance of being odd}

\author{Emmanuel Trizac}
\affiliation{LPTMS, CNRS, Univ. Paris-Sud, Universit\'e Paris-Saclay,
91405 Orsay, France}

\author{Gabriel T\'ellez}
\affiliation{Departamento de F\'{\i}sica, Universidad de los Andes,
  Bogot\'a, Colombia}

\begin{abstract}
From cement cohesion to DNA condensation, a proper statistical physics treatment of systems with long range forces is important 
for a number of applications in physics, chemistry, and biology. We compute here the
effective force between fixed charged macromolecules, screened by oppositely charged mobile
ions (counterions). We treat the problem in a one dimensional configuration, that allows for
interesting discussion and derivation of exact results, remaining at a level
of mathematical difficulty compatible with an undergraduate course. Emphasis
is put on the counter-intuitive but fundamental phenomenon of like-charge attraction,
\emma{that our treatment brings for the first time to the level of undergraduate 
teaching}. 
The parity of the number of counterions is shown to play a prominent role, which
sheds light on the binding mechanism at work when like-charge macromolecules do attract.
\end{abstract}

\maketitle

\section{Introduction}

\emma{Soft matter refers to a wealth of systems that may seem disparate at first sight:
foams, glues, paints, liquid crystals, colloids, polymers etc. \cite{Jones}. What they have 
in common is threefold: a) they are sensitive to `gentle' mechanical of thermal fluctuations;
b) they feature at least two distinct lengthscales (one being microscopic,
the other being intermediate between microscopic and macroscopic; c) they often exhibit strong collective effects,
such as the property to self assemble into complex entities. As a consequence, 
their properties are often difficult to rationalize and the sole knowledge of the basic
interaction forces between the individual components is not enough to understand 
the resulting properties. Here, our interest goes to one such collective effect,
that challenges intuition and takes place when charges are considered. It is at the root 
of key phenomena, outlined below. Soft Matter is in itself a 
relatively new field, but it is starting to enter the classroom 
\cite{class0,class1,class2,class3,class4,class5,class6}.
}

In vacuum, two like-charges interact {\em via} Coulomb law, which yields a repulsive force.
In a solvent like water, containing
mobile cations and anions, the situation becomes more complex. 
It is highly relevant in soft matter physics \cite{Jones}, and more precisely for
colloidal suspensions, where large charged macromolecules, typically of micron-size,
are surrounded by much smaller ions, called microions. As microions move on a
much shorter time scale than the macromolecules, the relevant concept to discuss interactions
is that of {\em effective potential}:
the direct interactions between the charged macromolecules are modified as a result
of the fluctuations of the medium. This yields an effective
interaction which is mediated by the microions, and it is obtained by
performing a thermal equilibrium average over the fluctuations of the positions of
the microions \cite{Chandler}. A first effect of the microions is to screen the bare
Coulomb repulsion, resulting in a short range effective interaction
between the macromolecules \cite{Levin}. In a small coupling regime (when the
typical electrostatic energy is smaller than the thermal one), the
effective interaction remains nevertheless repulsive between like
charges. However at large couplings, counterintuitive effects
appear such as attraction between like-charge bodies.

The phenomenon of like-charge attraction and its rationale are an old controversy
in physico-chemistry literature. While it surfaced as early as the 1930s \cite{LeDu39},
it appears that early accounts where flawed (as e.g. pointed out in \cite{VeOv48,Rque1}).
A breakthrough was realized in the 1980s when numerical evidences demonstrated 
that equivalently charged surfaces could attract each other, at short distances and under
high enough Coulombic coupling \cite{GJWL84,Note1,Note2}. It is now realized that like-charge 
attraction is the root of a wealth of phenomena in man-made or natural systems,
among which the formation of DNA condensates \cite{Bloom96,GBPP00}, or the cohesion of 
cement \cite{PevD04,IKLF16,Rque2}. \emma{We refer in particular the reader to \cite{GBPP00}
for an interesting and non technical introduction to these questions pertaining to
electrostatics.}

\emma{
The motivation for our paper is that like-charge attraction, in spite 
of its importance, is absent from educatinal textbooks. The reason is that
it is a complex phenomenon, and that analytical treatments are difficult.
On the other hand, computational studies are highly non-trivial and thus not
an educational option,
since one first has to learn how to deal with Coulomb forces, and more generally,
long range interactions \cite{AT}}.
In this article, we present a model where like-charge attraction is put in
evidence, \emma{in terms that are accessible to an undergraduate student
having only basic notions of electrostatcs, and of statistical physics}. 
To this end, a simplified model in one dimension will be worked out, 
which has
the advantage to be tractable analytically.
In real applications, in three dimensions,
the situation is more involved. Nevertheless, the model presented here not only
gives a simple introduction to the basic mechanisms behind like-charge attraction 
in colloidal systems, but also allows to recover exact results, as we shall emphasize 
below. \emma{In this sense, our treatment is true to experimental reality}.
The outline of the paper is as follows. The model will be presented in section 
\ref{sec:model}. For the sake of simplicity, we will address the situation where
one type only of microion is present and neutralizes the charge 
of two macromolecules. In such a case, the microions are referred to as
counterions, since their charge is opposite in sign to that of the macromolecule.
While this is a simplification, it is important to stress that certain 
experimental conditions come close to this so-called {\em de-ionized} limit
\cite{PMGE04}. After having defined the model, we will turn in section \ref{sec:one}
to the simplest situation where the system has only $N=1$ counter-ion.
Comparison to the $N=2$ case in section \ref{sec:two} will reveal that 
the parity of $N$ plays a fundamental role. This feature  will be elucidated in 
section \ref{sec:odd}, and highlights the fact that odd values of $N$ only can lead 
to attractive interactions. Finally, section \ref{sec:many} will be for 
the relevant large $N$ limit and section \ref{sec:concl} for our concluding remarks.

\section{The model}
\label{sec:model}

\subsection{From 3D to a 1D setting}

Consider two charged parallel plates. These will mimic the surface of
two colloidal charged particles (the ``macromolecules''). Between them, there is a given
number of counterions of opposite charge that screen the macromolecules. 
The system is globally electro-neutral, and considered in thermal equilibrium at a
temperature $T$.

We will be interested in computing the effective
force between the plates after averaging over the microscopic
configurations of the microions \cite{rque2017}. This can be obtained by several
means. One way is through the derivative
of the free energy of the system with respect to the separation of the
plates. The free energy can be computed from the canonical
partition function \cite{Chandler}. An alternative 
but of course equivalent route is through the use of a
relation between the density of counterions at the plates and the
pressure, which is known as the contact theorem~\cite{contact}, 
to be explained later on.

Due to the translational symmetry along the planes of the charged
plates, the density profile of the counterions depends only on the
coordinate perpendicular to the plates. Let us call this direction
$x$, and locate the first plate at $x=0$ and the second one at $x=L$
where $L$ is the distance between the plates. A further simplification
that will allow us to obtain analytical results is the smear out of the
charge of each counterions in a plane parallel to the charged
plates. With this ({\it a priori} unphysical but as will appear below fruitful)
approximation the system becomes effectively one
dimensional. The interaction between the ``smeared out''
counterions depends only on the $x$ coordinate and it is given by the
one-dimensional Coulomb potential. At this point, it is instructive to
review some basic facts about electrostatics in 1D.

\subsection{One dimensional specifics}

It is well-known that the electric field created by a charged infinite
plane is perpendicular to this plane, proportional to its surface
charge, and independent of the distance to the plane. The
associated electrostatic potential $v(x)$ at a distance $x$ from
the plane is \cite{Jackson}
\begin{equation}
  \label{eq:pot-1d-SI}
  v(x)= - \frac{\sigma}{2\epsilon_0} |x|  
  \,,
\end{equation}
where $\sigma$ is the surface charge density of the plane and
$\epsilon_0$ is the permittivity of vacuum. It is a solution of
Poisson equation in one dimension
\begin{equation}
  \label{eq:Poisson-1d-SI}
  \frac{d^2v}{dx^2}=-\frac{\sigma}{\epsilon_0}\, \delta(x)
  \,,
\end{equation}
where the Dirac distribution reflects the presence of the source
of the electric field at $x=0$.
With the model presented in the previous section, we consider a
collection of parallel charged planes interacting through the
potential~(\ref{eq:pot-1d-SI}). Equivalently one can think of a
one-dimensional world where point charges live on a line and
interact with the same potential. In this one-dimensional view, it is
more natural to work in units where the potential is
\begin{equation} 
  \label{eq:pot-1d}
  v(x)= - q |x|  
  \,,
\end{equation}
and $q$ will be called the charge of the particle. The corresponding
electric field is simply $\mathbf{E}=-v'(x) \hat{\mathbf{x}}$ with
\begin{equation}
  \label{eq:E-1d}
  -v'(x) = 
  \begin{cases}
    + q & \text{for positions at the right side of the 
      charge ($x>0$)\,,}\\
    - q & \text{for positions at the left side of the charge ($x<0$)\,.}
  \end{cases}
\end{equation}
Consider now a collection of charged particles on the line. One can
obtain the electric field at a given position by applying
Gauss law and the superposition principle \cite{Jackson}. The result is simple:
the electric field is given by the difference of the sum of the
charges at the left and the right sides of the considered
position. This field remains constant as long as one does not cross a
particle, in which case it will change by an amount $\pm 2 q$, where
$q$ is the charge that has been crossed, and the sign depends on
whether the cross was from left to right ($+$) or right to left ($-$).

In our one-dimensional picture, the
situation is that of two charges $Q>0$ located at $x=0$ and $x=L$. Between
them, there are $N$ counterions, which are particles with
charge $e=-2Q/N$ at positions $x_1, \cdots, x_N$. Overall, the system
is thus neutral. The charges $Q$ are
at fixed positions, but the counterions can freely move, and are at
thermal equilibrium at temperature $T$ \cite{Chandler}. The natural framework to
study this system is the canonical ensemble \cite{Chandler}. Let us denote $Z(N,L,T)$
the canonical partition function of the system. The corresponding free energy 
is $F(N,L,T)=-k_B T \ln Z(N,L,T)$ where $k_B$ is Boltzmann
constant. The effective force between the (macromolecular) charges $Q$ located at $x=0$
and $x=L$ is $\mathbf{F}= P(N,L,T) \hat{\mathbf{x}}$ where $\hat{\mathbf{x}}$
is a unit-vector along the $x$-axis and
\begin{equation}
  \label{eq:P-def}
  P(N,L,T) \, =\, -\frac{\partial F(N,L,T)}{\partial L} \,= \,
  k_B T \,\frac{1}{Z}\frac{\partial
    Z}{\partial L}
  \,.
\end{equation}
Before considering some particular cases for a few values of $N$,
it is instructive to state the contact theorem, that provides a
sometimes useful alternative to the force calculation along the lines
of Eq. \eqref{eq:P-def}.

\subsection{Contact theorem}

Consider an arbitrary ensemble of charges confined between two parallel plates
(charges on a segment between two points as in our 1D setting,
or charges in a slab between two parallel lines in 2D, or charges between parallel 
planes etc). We denote by $\rho(x)$ the equilibrium average density of
ions, that depends on position $x$.  The contact theorem~\cite{contact},
states that the force $P$ and
the density $\rho$ are related by
\begin{equation}
  \label{eq:contact}
  P \, = \, k_B T \rho(0) - Q^2
  \,.
\end{equation}
The density $\rho(0)$ is evaluated at contact with the confining boundary at
$x=0$. Each contribution can be understood as
follows. The term $k_B T \rho(0)$ stems from the force due
to the collisions of the counterions against the wall at $x=0$. It takes an ideal gas 
form here \cite{rque20}, and is always repulsive.
The second term, $-Q^2$, is attractive since global neutrality 
requires that the total charge, excluding that of the ``macromolecule'' ($Q$), is 
exactly \gtad{the opposite of $Q$}. This electrostatic
contribution to the force is more familiar when written in terms of more familiar quantities
such as $\epsilon_0$ \cite{rque30}; it is known as the electrostatic pressure,
of paramount importance in the physics of conductors \cite{Jackson}.

It will turn useful for our purposes to generalize the contact theorem
to the case of confinement by asymmetric boundaries, with charge $Q_1$
at $x=0$ and charge $Q_2$ at $x=L$. The force can indifferently be
computed \gtad{using the densities and charges at either boundary},
with the result
\begin{equation}
  \label{eq:contact_general}
  P \,=\, k_B T \rho(0) - Q_1^2 \,=\,  k_B T \rho(L) - Q_2^2
  \,.
\end{equation}
This will be used in section \ref{sec:odd}, in a situation where asymmetry does not come
from $Q_1\neq Q_2$ but from the presence of counterions of different charges. 
Note that the difficulty 
with the contact theorem is that in general, the contact densities 
$\rho(0)$ and $\rho(L)$
are not known. Yet, Eq. \eqref{eq:contact_general} amounts to a 
remarkable connection between them, that should always 
hold.

\section{One counterion $N=1$.}
\label{sec:one}

Let first consider the simplest case when there is only one counterion
($N=1$) located at a position $x_1\in[0,L]$. For the system to be
globally neutral, we require the counterion charge to be $-2Q$,
thereby balancing the two charges $Q$. In the region $[0,L]$ where the
counterion is confined, the electric field created by the charge $Q$
at $x=0$ is cancelled by the field due to the other charge $Q$ at
$x=L$. Thus the counterion does not feel any electric force and it is
free to move in $[0,L]$.  The electric field felt by the charge $Q$
located at $x=0$ is obtained simply by summing the charges that are
located at its right: the counterion at position $x_1$ (charge $-2Q$)
and, at $L$, the charge $Q$. In total, the electric field is
$-(-2Q+Q)\hat{\mathbf{x}}=Q\hat{\mathbf{x}}$ \gtad{and the force on the
charge $Q$ located at $x=0$ is $Q^2\hat{\mathbf{x}}$}. By the same
argument, the electric field felt by the charge $Q$ located at $x=L$
is $-Q\hat{\mathbf{x}}$ \gtad{and the corresponding force is
$-Q^2\hat{\mathbf{x}}$. These three forces derive from the potential energy }
\begin{equation}
  \label{eq:U1}
  U=Q^2 L
  \,.
\end{equation}
\gtad{It can be easily checked that the force on each particle are
  reproduced computing minus the gradient of the potential energy with
  respect to each of the particles positions.}
  
\begin{figure}
  \begin{center}
    \includegraphics[width=0.6\textwidth]{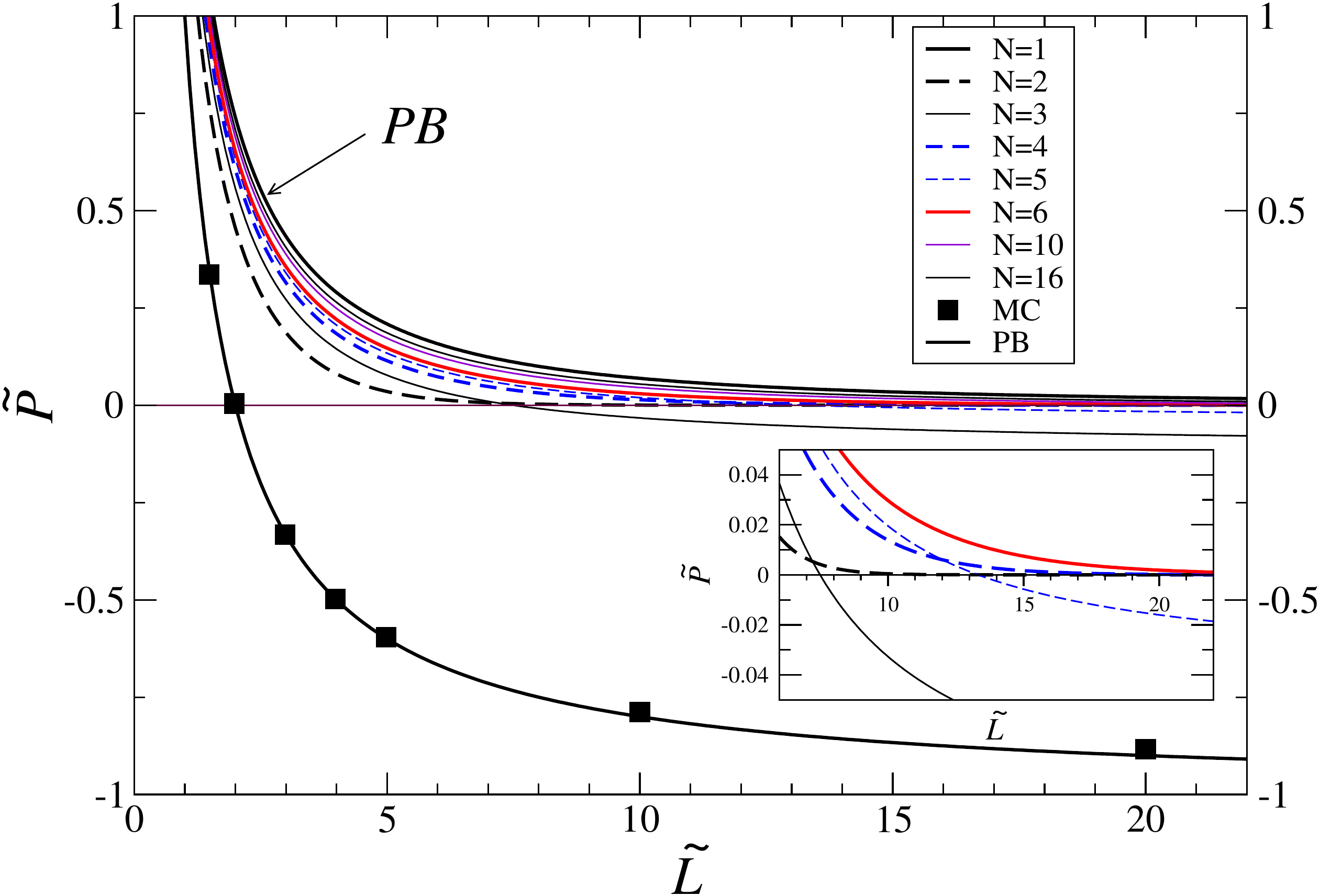}
  \end{center}
  \caption{Force $\widetilde{P}=P/Q^2$ as a function of the charges
    separation $\widetilde{L}=2 L Q^2/(k_B T)$, for different number of counterions. 
    Close inspection reveals that odd $N$ only can lead to attractive interactions
    ($\widetilde P<0$).
    The results are compared to Monte Carlo simulation data of a {\it bona fide}
    three-dimensional system where counterions are confined by uniformly charged
    parallel plates, in a situation where Coulombic coupling is large 
    (square symbols, data from Ref \cite{MoNe02}).
    The curve denoted PB is for the result of Poisson-Boltzmann theory, where the
    calculation of the force is different in spirit, and 
    follows from the solution of a non-linear differential equation. \gtad{The
    inset provides a zoom into the region where some pressure curves become negative 
    (cases $N=3$ and $N=5$), while the even $N$ cases remain positive for all distances.}
  \label{fig:PvsL}}
  
\end{figure}

The partition function of the system is 
\begin{equation}
  \label{eq:Z1def}
  Z(1,L,T)=\frac{1}{\lambda}\int_0^{L} e^{-\beta Q^2 L} \,dx_1
  \,,
\end{equation}
where $\beta=1/(k_B T)$ and $\lambda=h/\sqrt{2\pi m k_B T}$ is the de
Broglie thermal wavelength of the counterion of mass $m$, and $h$ is
Planck constant. This quantum mechanics intrusion in the problem is immaterial,
and it will be important to check {\it a posteriori} that it does not 
affect physical observable like the force $P$.
Since $U$ does not depend on $x_1$, the calculation of the integral in \eqref{eq:U1} is
straightforward
\begin{equation}
  \label{eq:Z1res}
  Z(1,L,T)=\frac{L}{\lambda} e^{-\beta Q^2 L}
\,.
\end{equation}
The effective force between the charge is then obtained applying
(\ref{eq:P-def})
\begin{equation}
  \label{eq:P1}
  P(1,L,T)= \frac{k_B T}{L} - Q^2
  \,.
\end{equation}
The contact theorem, Eq \eqref{eq:contact}, would have readily provided us with the very 
same information,
since the counterionic density is $x$-invariant in our case (no electric field acting
on the counterion): thus, $\rho(x)$ is uniform 
in $[0,L]$, and normalization tells us that $\rho(x)=k_B T/L$.
This force, as expected, \gtad{is} independent of Planck constant, 
is plotted as a function of $L$ in Figure
\ref{fig:PvsL}. Note that for short separations $L$, it is repulsive
($P>0$) but as $L$ increases, the force decreases and eventually
becomes attractive ($P<0$) for $L>L^{*}=\beta Q^2$. This is a simple
illustration of the counterintuitive phenomenon of like-charge
attraction. The asymptotic behaviors for short and large distances are
\begin{equation}
  \label{eq:P1L0}
  P(1,L,T)\sim  \frac{k_BT}{L}\,, \quad L\to 0
\,,
\end{equation}
and
\begin{equation}
  \label{eq:P1Linf}
  P(1,L,T)\to - Q^2\,,
  \quad L\to\infty
  \,.
\end{equation}

In Figure~\ref{fig:PvsL}, we show Monte Carlo (MC)
simulations results for the effective force between two parallel charged
plates sandwiching a large number of counterions, in the strong coupling
regime \cite{MoNe02,Rque15}. Note how accurately this simple one-dimensional model captures
the behavior of the more realistic situation in three dimension.
Thus, although the MC 3D simulations are run with a large value of $N$,
the distance range corresponding to Fig. \ref{fig:PvsL} is such that 
the typical inter-counterion distance is much larger than $L$,
the inter-plate distance. Then, counterions become decoupled and the physics
is equivalent to that of our $N=1$ counterion in a slab \cite{DHNP09,TT2015}. We emphasize 
that upon 
further increasing the distance range in the figure (so as to exceed the typical distance
between counterions), the MC symbols
would eventually reach the line $\widetilde P=0$. For the 
corresponding very large distances, our one-dimensional model ultimately becomes irrelevant.
The fact that we find that $\widetilde P$ does not vanish for $L\to\infty$ 
is a peculiarity of one-dimensional electrostatics, which does not transpose
to higher dimensions.

\section{Two counterions $N=2$.}
\label{sec:two}

Let us now turn our attention to the case where there are two
counterions, each of them with a charge $-Q$, at positions $x_1$ and
$x_2$. Without loss of generality, we order the particles
positions as $0<x_1<x_2<L$. The electric field at $x=0$ and $x=L$ is
the same as in the previous section. The electric field at the
position $x_1$ of the first counterion is obtained by counting the
charges at its left ($Q$) and its right ($-Q+Q=0$). The electric field
is then $Q \hat{\mathbf{x}}$. For the particle at $x_2$ it is simply the
opposite. From this, we obtain the potential energy of the system
  \begin{equation}
    \label{eq:U2}
    U=-Q^2 |x_2-x_1| +Q^2 L 
    \,.
  \end{equation}
The partition function reads
\begin{equation}
  \label{eq:Z2def}
  Z(2,L,T)=\frac{1}{\lambda^2}
  \iint_{0<x_1<x_2<L} e^{\beta Q^2
    (x_2-x_1 - L)} \,dx_1\,dx_2
  \,.  
\end{equation}
Once the integrals are computed, the result can be cast as 
\begin{equation}
  \label{eq:Z2res}
  Z(2,L,T)=\frac{
    1-e^{-\beta Q^2 L} \left(1+\beta Q^2 L\right)
  }{(\beta Q^2\lambda)^2} 
  \,.
\end{equation}
The effective force is then
\begin{equation}
  \label{eq:P2}
  P(2,L,T)= \frac{\beta Q^4 L}{e^{\beta Q^2L} - 1 - \beta Q^2 L }
  \,,
\end{equation}
once more independent of the de Broglie wavelength $\lambda$, and thus of Planck's constant,
as it should.  
Since the exponential is a convex-up function, we have $e^x>1+x$ for all $x$, so that now
$P(2,L,T)>0$ for all values of $L$: the force is repulsive. At short distances
\begin{equation}
  \label{eq:P2L0}
  P(2,L,T)\sim  \frac{2 k_BT}{L}\,, \quad L\to 0
\,,
\end{equation}
which is readily recovered from the contact theorem \eqref{eq:contact}. Indeed, for strong confinement
(small $L$), the repulsive contribution in $\rho(0) k_B T$ is dominant since the 
counterion density $\rho$ diverges, the latter being given by $2/L$.
For large distances, the force decays exponentially
\begin{equation}
  \label{eq:P2Linf}
  P(2,L,T)\sim \beta Q^4 L e^{-\beta Q^2 L}\,,
  \quad L\to\infty
  \,.
\end{equation}
This exponential form is a fingerprint of complete screening of the macromolecule
charges by the counterions.

For $N=2$, like-charge attraction is thus ruled out.
What
is the physical mechanism behind like-charge
attraction with $N=1$ counterion, but discarded with $N=2$? An important
difference between these two cases is the ``sharing'' of a counterion
between the fixed macromolecules $Q$ at $x=0$ and $x=L$. For $N=1$, the
counterion does not feel any electric field, as the field created by
both macromolecules cancels out.  The counterion thus
does not have any preference and can roam in the full region between
the charges $Q$. This counterion shared by the charges creates a bond
between them, that in turn, leads to 
attractive interactions at large enough distance. Note that attraction cannot
hold at short separation, since as noticed above, the repulsive (entropic)
contribution embodied in the contact density of counterions $\rho(0)$ 
becomes divergent, with a resulting dominant repulsive squeezing effect.

When $N=2$, on the contrary, the left ion at $x_1$ 
feels an electric field that pushes it to
the left-most macromolecule $Q$ located at $x=0$. Similarly, the counterion on the
right hand-side (position $x_2$) is subject to an electric field that pushes
it to the right-most charge $Q$ at $x=L$. At large enough distance $L$,
the resulting picture is
that of two {\em decoupled} neutral objects without the possibility of presenting an
effective attractive force. We can surmise at this point
that a counterion can be shared whenever $N$ is an odd number, with a
foreseeable attraction, while the sharing (or coupling) effect will disappear
whenever $N$ is even. In a nutshell, we expect to observe like-charge attraction
for odd $N$, and none for even $N$. This is indeed correct \cite{TT2015}.
In the following section, we focus on the situation where the number
of counterions $N$ is arbitrary but odd, in an effort to
compute exactly the large distance force between the macromolecules.

\section{An odd number of counterions $N=2n+1$.}
\label{sec:odd}

\begin{figure}
  \begin{center}
    \includegraphics[width=0.7\textwidth,clip]{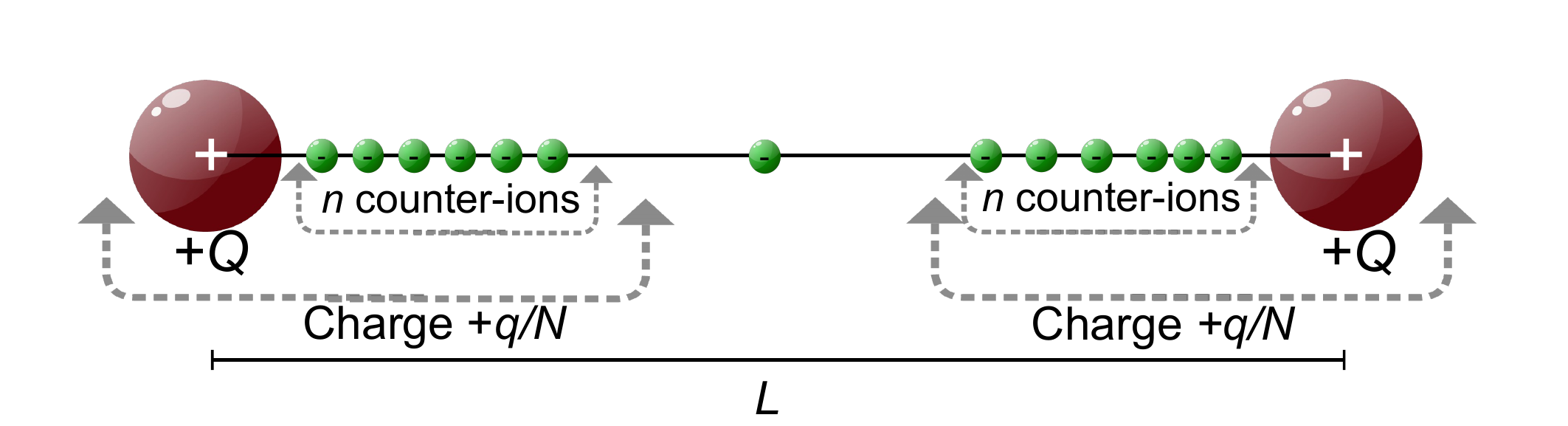}
  \end{center}
  \caption{Screening of two charges by an odd
    number $N=2n+1$ of counterions, where $n$ is an integer. 
    The fixed ``macromolecules'' both have charge $+Q$. 
    For the sake of illustration, they are shown as spheres,
    but in the present one-dimensional setting, they simply are point-like,
    located at $x=0$ and $x=L$. The configuration displayed is relevant at large 
    $L$, where the ``misfit'' central counterion can be seen. It acts as a bridge 
    between its right and left-hand sides, that are thereby electrostatically bound
    and coupled.
    This misfit ion, by symmetry free of electric force, 
    is the mediator of like-charge attraction, see text.
  \label{fig:odd-counterions}}
  
\end{figure}

For arbitrary $N$, the computation of the partition function becomes
involved, although it can be performed
explicitly~\cite{Len61,Pra61,DHNP09,TT2015}. However, it is not
necessary to do the explicit calculation to grab interesting characteristics
of the effective force: a ``trick'', that we now present, 
allows to find  the exact force at large distances.
The argument goes in two steps and we start by a qualitative discussion.
If $N=2n+1$, $n$ being an integer, global neutrality demands that each counterion bears 
a charge $-2Q/(2n+1)$. The first $n$
counterions at the left hand side of the system are in an electric
field oriented to the left and they will be attracted to the charge
$Q$ located at $x=0$. This creates a charged object with charge
$Q-n\frac{2Q}{2n+1}=Q/(2n+1)$. Similarly, the $n$ counterions at the
rightmost region will be attracted to the charge $Q$ located at $x=L$
creating an object with the same charge $Q/(2n+1)$. In the middle,
there remains a ``misfit'' counterion, see
Figure~\ref{fig:odd-counterions}. Since its right and left neighbors bear the same
total charge, the electric field in the corresponding region is zero, by symmetry. 
Apart from confinement effects (that will become less and less important as $L$ increases),
the misfit counterion is free to roam, without any electric
force acting on it. The situation is then reminiscent of that when 
$N=1$. This counterion will create a bond between the two
charged objects, which will result in an effective attraction if $L$
is large enough. Let us show this in a more quantitative way.

We now turn to the second step of our argument, that relies 
first on the definition of an auxiliary useful and asymmetric problem, followed 
by an application of the contact theorem in that asymmetric setting.
We take $L$ large, meaning more precisely that 
$L\gg 1/(\beta Q^2)$. To compute
the density, we consider an auxiliary model where we regroup the
leftmost $n+1$ counterions into a single composite ion of charge $-(n+1)
(2Q/N)=-(Q+Q/N)$. For large $L$, we argue that this does not modify the force on the
charge $Q$ located at $L$ (decreasing $L$, the statement becomes less and less correct). 
The composite ion is in an electric field
due to the charge $Q$ at the left ($x=0$) and a charge $-n (2Q/N) + Q
= Q/N$ at the right. This electric field is $Q-Q/N$. Denoting $x$ the position of
the composite ion, the corresponding potential energy
stems from the product of the composite ion charge with the electric field:
$-(Q+Q/N)(Q-Q/N) x= Q^2 (1-(1/N)^2) x$. The composite ion density
then follows from the Boltzmann factor associated to this
energy:
\begin{equation}
  \label{eq:rho1}
  \rho(x) = C e^{-\beta Q^2 (1-\frac{1}{N^2})x}
  \,.
\end{equation}
The normalization constant $C$ is obtained by requiring that $\int_0^{\infty} \rho(x)\,dx=1$. 
Strictly speaking, the upper bound of integration should not be $\infty$, because 
we always are in a finite $L$ system. Yet, the error incurring becomes exponentially small in
$L$, and we finally arrive at
\begin{equation}
  \label{eq:rho2}
  \rho(x) = \beta Q^2\left(1-\frac{1}{N^2}\right) 
  e^{-\beta Q^2 (1-\frac{1}{N^2})x}
  \,.
\end{equation}
From this, we obtain the contact density $\rho(0)=\beta Q^2\left(1-\frac{1}{N^2}\right)$.
Invoking one last time the contact theorem, in its asymmetric formulation
embodied in Eq. \eqref{eq:contact_general}, we get the force as 
\begin{equation}
  \label{eq:PLinfty}
  P=-\frac{Q^2}{N^2}\,, \qquad L\to\infty\,.
\end{equation}
An attractive interaction ensues for large $L$. This is the
generalization of the result derived for $N=1$, to any odd $N$. Figure~\ref{fig:PvsL} shows the
pressure as a function of $L$ for different $N$-values. There, it
can be noticed that for $N$ even, the force is always repulsive ($P>0$)
while on the contrary, when $N$ is odd, the force becomes attractive
for a sufficiently large inter-charge distance $L$.

\section{Large $N$ limit.}
\label{sec:many}

In the odd number of counter-ion case, where like-charge attraction is 
possible, the amplitude of the effect decreases as $1/N^2$ when $N$ increases,
and practically vanishes whenever $N$ exceeds a few tens. 
This is illustrated in Figure~\ref{fig:PvsL}. In the $N\to \infty$
limit, the charge of the ions $e=-2Q/N$ vanishes, but there is a large
number of them. An adequate description of
the system is to consider that the counterions form a continuous
charge density. \gtad{Furthermore, since in that limit 
  the electrostatic coupling between the ions vanishes ($e=-2Q/N\to
  0$),} this situation is described by a so-called 
mean-field theory, where the field in question refers to the
electrostatic potential \cite{rque123}. The charge density can be obtained by solving
Poisson--Boltzmann equation~\cite{PB,Rque10}. It has been shown
that within such a framework, like-charge attraction is 
impossible~\cite{PBrepulsion}, as it has been illustrated here. The 
attraction under scrutiny in the previous section was possible due to
the ``sharing'' of one counterion (the misfit). The ``granularity'' of matter, which
is important for this phenomenon to happen, is lost in the mean-field
description. An equivalent formulation is to note that replacing 
a finite number of particles by a field, as useful as it can be in some cases,
discards an important length scale related to inter-ionic typical distances.
This length plays a  key role in the like-charge attraction phenomenon 
\cite{SaTr12}. It is precisely the typical distance
between counterions, featured at the end of section \ref{sec:one}.

\section{Discussion and Conclusion}
\label{sec:concl}

We have studied a simple model of point charges on a line, that mimics 
the situation of two charged macromolecules of charge $Q$, held fixed in space at a distance $L$, and
screened by a collection of microions. This is a
prototypical soft matter question, of paramount importance in various fields
of colloid science, from physics to biochemistry, including food industry.
Compared to experimental systems, we made the simplifying 
assumption that microions are only of one type (the counterions).
In reality, microions of both signs are usually present (e.g. Na$^+$ and Cl$^-$),
but it is important to stress that the counterions-only limiting case 
can be approached experimentally, through a process called deionization.

In our presentation, the contact theorem sketched in section \ref{sec:model}
plays a particular role. 
This exact statistical mechanics relation expresses the force felt by one
macromolecule as a tradeoff between the repulsive kinetic pressure
due to the collisions with counterions (local interactions),
and the attractive electrostatic pressure (long-range, non-local interactions).
We started in section \ref{sec:one} with the extreme case where only $N=1$ counterion screens the
two macromolecules. We could explicitely check that the contact theorem routes yields
the same force between the macromolecules, as the route going through the calculation 
of the partition function, with subsequent $L$-derivative. Interestingly,
these results allow to recover those of 3D numerical simulations,
provided two requirements are satisfied: one should be in a regime where the
macromolecule charge is large enough (strong coupling condition), 
and in a distance range where the
typical 3D distance between counterions is larger than $L$ (see the agreement between our $N=1$ 
curve in Fig. \ref{fig:PvsL} and the exact Monte-Carlo data).

Going from the $N=1$ case to $N=2$ in section \ref{sec:two}, we noticed that the effective 
force turns from log-range attractive to all distance repulsive. 
This pinpoints the significance of the parity of $N$. 
For odd $N$, we indeed showed in section \ref{sec:odd},
invoking a trick where some carefully chosen 
counterions are grouped to form a composite ion 
(so as not to affect the force we aimed at computing), that
the large $L$ force is attractive, given by $-Q^2/N^2$. 
The rationale behind attraction is that for odd $N$, there always exists a ``misfit''
counterion, electric-field free by symmetry, that floats in between its neighbors, 
and that becomes completely delocalized when $L$ increases. This delocalized ion
has a bridging effect that couples the two moieties of the system; 
hence the negative force at large 
$L$. On the other hand, no ion can become delocalized when $N$ is even,
and effective interactions are always repulsive. At short distances and for all
$N$, the force is always repulsive, as a squeezing penalty due to 
the entropic cost for confining ions, that becomes overwhelming. 

We believe that these considerations, relying on simple algebra, can be 
useful for an undergraduate course in statistical physics. Such a discipline
is difficult to marry with long-range interactions like those in Coulomb
systems. In particular, our final section (\ref{sec:concl}) on the large $N$ limit 
where the relevant approach is no longer particle-based, but field-theoretic,
provides interesting additional pedagogical perspectives for a more advanced and comprehensive discussion,
still at a level of mathematical complexity that remains affordable.

\begin{acknowledgments}
G.T. acknowledges support from Fondo de Investigaciones, Facultad de
Ciencias, Universidad de los Andes, Project 2017 ``Efectos de apantallamiento fuerte en modelos simples''.
\end{acknowledgments}


\begin{thebibliography}{99}
\bibitem{Rque2}
In cement, calcium silicate hydrates are an essential constituent. These hydrates 
have a debated structure and microstructure, but they can schematically be viewed as forming
large and to first approximation planar layers, sandwiching divalent microions. The cohesion 
between neighboring layers is to a large extent due to like-charge attraction.
\bibitem{rque2017}
\emma{
In an experimental system, solvent molecules would also be present. 
They are discarded from our discussion, since in the simplest approximation,
a solvent like water can be treated as a dielectic and structureless continuum.
What matters is that the dielectric permittivity of the solvent is not the same
as that of vaccuum, but apart from the change in the main text $\epsilon_0 \to \epsilon$,
the solvent can be left aside.
}
\bibitem{rque30}
It may useful to rewrite the contact theorem in terms of more standard quantities
pertaining to usual three-dimensional electrostatics. The pressure at a flat 
interface having uniform surface charge $\sigma$ then reads
$P = k_B T \rho(0) - \sigma^2/(2\epsilon_0)$, where $\epsilon_0$ is the 
permittivity of vacuum. The (attractive) contribution in $\sigma^2$ is the so-called
electrostatic pressure, see e.g. \cite{Jackson}, chapter 1, section 1.11.
\bibitem{Rque15}
For the Monte Carlo simulations shown, the coupling parameter defined in \cite{MoNe02}
is $\Xi=10^5$. It is large enough to qualify the system as ``strongly coupled''.
\bibitem{rque123}
\gtad{A more rigorous analysis can be found in section IV of Ref. 
\cite{DHNP09}, where it is shown that the
relevant coupling parameter, $\Xi$, behaves in the present 1D problem like 
$1/N$. Hence, the large $N$ limit is that where mean-field theory becomes exact. 
This is at variance with the three dimensional situation.
}
\end{thebibliography}
\end{document}